\documentclass[5p,twocolumn]{elsarticle} 
\usepackage{graphicx}
\usepackage{epstopdf}

\begin{document}

\oddsidemargin = -12pt

\title{On the symmetries of the $^{12}$C nucleus}

\author{J. Cseh  and R. Trencs\'enyi\\
Institute for Nuclear Research, Hungarian Academy of Sciences,
                Debrecen, Pf. 51, Hungary-4001} 
 
\date{\today}

\begin{abstract}
The consequences of some symmetries of the three-alpha system
are discussed. In particular, the recent description of the
low-energy spectrum of the
$^{12}$C nucleus
in terms of the Algebraic Cluster Model (ACM)
is compared to that of the Semimicroscopic Algebraic Cluster Model (SACM).
The previous one applies interactions of a  $D_{3h}$ geometric symmetry 
\cite{evid},
while the latter one has a U(3) multichannel dynamical symmetry,
that connects the shell and cluster pictures. 
%The permutational symmetry of the three alphas are involved in 
%both of them, and the SACM involves the Pauli-principle, too.
The available data is in line with both  descriptions.
\end{abstract}

\begin{keyword}
alpha-clustering, geometrical, dynamical, and permutational symmetries
\PACS 21.60.Fw, 21.60.Cs, 27.20.+h
\end{keyword}

\maketitle

\oddsidemargin = -12pt

The $^{12}$C nucleus plays a crucial role in the evolution of stars,
it is an important test ground of different models of atomic nuclei,
furthermore, it is a rich laboratory of symmetries.
Therefore, it is in the focus of both the experimental and
the theoretical research. New states or transitions detected in
this nucleus can choose between competitive models.

The recent observation of the 5$^-$ state by the Birmingham 
group 
\cite{evid}
%inspired promtpt theoretical activity, and it 
was interpreted as
an evidence for the $D_{3h}$ symmetry in nuclear structure.
%\cite{evid}.
The algebraic  cluster model
\cite{bipr,biap}
describes the detailed rotational-vibrational spectrum of the three-alpha
system, and the application  of its $D_{3h}$ limit  results in an energy spectrum
that is in good agreement with the experimental observation
\cite{evid}.

Alpha-cluster models have been applied to the $^{12}$C very extensively
\cite{exte},
from the beginning of nuclear structure studies 
\cite{begi}
until very recently
\cite{rece}.
It seems to be a general agreement of these cluster studies that the
ground state of the $^{12}$C nucleus has a triangular shape at the vertices
with the three alpha-particles. The novel feature of the algebraic cluster model
interpretation 
\cite{evid,bipr,biap}
is twofold. First: in this model not only the ground state, or the ground
band is considered to have the $D_{3h}$ symmetry, but the excitation
spectrum, too, including the Hoyle-band, and several others.
Second: the algebraic cluster model describes in detail the spectrum 
of the three-alpha system with the triangular symmetry, as opposed to 
e.g. the Bloch--Brink alpha-cluster model, which gives this shape for
the ground state, but it does not provide us with detailed spectrum
\cite{blbr}. 

In Table 1 we show the symmetries of some alpha-cluster models 
applied to the $^{12}$C nucleus, including the present work.
The interactions applied in the calculations are also listed.
The microscopic cluster model (MCM) in the last line actually
indicates a family of models (GCM, RGM, OCM...). From the viewpoint 
of the applied symmetries they are similar to each other. 
For their detailed discussion and comparison we refer to
\cite{hoik}. 

A new chapter of the structure studies of  $^{12}$C is the application of those
microscopic calculations, which describe each nucleonic
degrees of freedom without supposing any cluster structure
\cite{amd,fumi,luis},
and in some cases realistic nucleon-nucleon forces are applied, 
i.e. real ab initio calculations are carried out
\cite{abin}. 
Obviously, for the understanding of the structure, these
approaches are the most promising.

In this paper we investigate the three-alpha cluster system from a
different angle. In particular, we compare the consequences of different
symmetries of the system. The $D_{3h}$ point symmetry, mentioned beforehand,
is a geometrical one. In addition to this, the $S_3$ permutational symmetry
of the three identical alpha-particles is also essential. It is involved both in the algebraic
and in the other cluster models. A further basic symmetry is the antisymmetry
of the 12 nucleons building up the $^{12}$C, resulting in  the Pauli-exclusion principle. 
This is taken into account in the microscopic (and semimicroscopic) cluster models, 
but it is not involved in
the phenomenologic ones, like the algebraic cluster model.
Here we make a comparison between the
performance of the algebraic cluster model with $D_{3h}$ symmetry and
that of a U(3) multichannel dynamical symmetry
\cite{mus1,mus2}. 
This latter one is the connecting symmetry
of the cluster and shell (quartet) models, which is formulated in  
the semimicroscopic algebraic approach
\cite{sacm,saqm}.

\begin{table}
%\vspace*{1mm}
\begin{tabular}{|c|c|c|c|c|c|}
\hline
\hline
%\vskip 4pt
Model&Interaction&$S_3$&$D_{3h}$&Pp&Spectrum\\ \hline
ACM&$D_{3h}$&$+$&$+$&$-$&+\\ \hline
SACM&U(3) MUSY&$+$&$-$&+&+\\ \hline
Bl-Br&$Volkov, Brink$&$+$&$-$&+&$-$\\ \hline
MCM&$various$&$+$&$-$&+&+\\ \hline
\hline
\end{tabular}
\caption{
The symmetries of some 3-alpha-models of the 
$^{12}$C nucleus. 
ACM: algebraic cluster model,
SACM: semimicroscopic algebraic cluster model,
Bl-Br: Bloch--Brink alpha-cluster model,
MCM: microscopic cluster model.
$S_3$ stands for the permutational symmetry of the three 
identical particles, 
$D_{3h}$ is the geometrical symmetry of the equilateral triangle,
while Pp means Pauli-principle. The last column indicates if detailed spectrum 
is provided by the model.
}
\end{table}

In what follows, first we introduce the semimicroscopic algebraic quartet (SAQM)
and cluster (SACM) models and their connecting multichannel dynamical symmetry
(MUSY). Then we present the U(3) MUSY spectrum in comparison with the
experimental one and with the  spectrum of the $D_{3h}$  model.
Finally, some conclusions are drawn.
\\

\noindent
{\it 
The  semimicroscopic algebraic quartet model} 
(SAQM)
\cite{saqm}
is a symmetry-governed truncation of the no-core shell model
\cite{nocore},
that describes the quartet excitations in a nucleus.
A quartet is formed by two protons and two neutrons,
which interact with each other very strongly,
as a consequence of the short-range attractive forces 
between the nucleons inside a nucleus
\cite{arima}.
The interaction between the different quartets is weaker.
In this approach 
the L-S coupling is applied, the model space has a spin-isospin sector
characterized by Wigner's U$^{ST}$(4) group
\cite{wigner},
and a space part described by Elliott's U(3) group
\cite{elliott}.
Four nucleons form a quartet  
\cite{harvey}
when their
spin-isospin symmetry is \{1,1,1,1\}, and their permutational symmetry is \{4\}. 
This definition allows two protons and two neutrons to form a
quartet even if they sit in different shells. As a consequence,
the quartet model space incorporates 0, 1, 2, 3, 4, ... major shell
excitations (in the language of the shell model),
contrary to the original interpretation of
\cite{arima},
when the four nucleons had to occupy the same single-particle
orbital, therefore, only 0, 4, 8, ... major shell excitations could be
described. 

The model is fully algebraic,
% in the sense that not only its basis states
%are characterized by group symmetries, but also 
%the physical operators are written in terms of group generators
%\cite{quart}.
therefore, group theoretical methods can be applied in calculating the matrix elements.
The operators contain parameters to fit to the experimental data,
that is why the model is called semimicroscopic: phenomenologic
operators are combined with microscopic model spaces.
Due to the quartet symmetry, only a single \{1,1,1,1\} U$^{ST}$(4) sector plays a 
role in the calculation of the physical quantities,
thus the U(3) space-group and its subgroups are sufficient for characterizing the situation:
%Eq.1
\begin{eqnarray}
U(3) \supset SU(3) \supset SO(3) \supset SO(2)
\nonumber\\
\vert [n_1,n_2,n_3]  ,  (\lambda , \mu) ,\  K\ ,  \ \ L \ \ \ \ ,\ \ \  M  \ \rangle .
\label{eq:ellgrch}
\end{eqnarray}
In Eq. (1) we have indicated also the representation labels of the groups,
which serve as quantum numbers of the basis states.  
Here $ n= n_1 + n_2 +n_3 $ is the number of the oscillator quanta, and
 $\lambda = n_1 - n_2, $
 $\mu = n_2 -n_3 $.
The angular momentum content of a $(\lambda ,\mu )$ representation is as
follows 
\cite {elliott}: 
$L= K, K + 1,...,K + max {(\lambda , \mu)}$, 
$ K = min {(\lambda , \mu )},
 min {( \lambda , \mu )}  - 2,..., 1 \ or \ 0,$
%\begin{equation}
 %\begin{array}{l}
 %\displaystyle{
 %L = K_L, K_L + 1,...,K_L + max {(\lambda , \mu)},
 %\label{mlett:1}
 %} \\  \ \\
 %\displaystyle{
 %K_L = min {(\lambda , \mu )},
 %min {( \lambda , \mu )}  - 2,..., 1 \ or \ 0,
 %} \end{array}
%\eqnum{4.2} 
%\end{equation}
with the exception of $K_L = 0$, for which
%\begin{equation}
$ L = max {(\lambda , \mu)}, max {(\lambda , \mu)} - 2,..., 1 \ or \ 0 $.
%\eqnum{4.3}
In the limiting case of the dynamical symmetry, when the Hamiltonian
is expressed in terms of the invariant operators of this group-chain,
an analytical solution is available for the energy-eigenvalue problem
(an example is shown below).

The SAQM can be considered as an effective model in the sense of
\cite{effective}:
the bands of different quadrupole shapes are described by their lowest-grade
U(3) irreducible representations
(irreps) without taking into account the giant-resonance excitations, built
upon them, and the model parameters are renormalized for the subspace of the
lowest U(3) irreps. 
\\

\noindent
{\it 
The semimicroscopic algebraic cluster model}
 (SACM) 
\cite{sacm},
just like  the other cluster models,
classifies the relevant degrees of freedom of the nucleus  into  two categories:
they belong either to the internal structure of the clusters, or to their relative motion.
In other words, the description is based on a molecule-like picture.
The internal structure of the clusters is handled in terms of
Elliott's shell model
\cite{elliott} with
U$^{ST}$(4)$\otimes$U(3) group structure
(as discussed beforehand).
The relative motion is taken care of by algebraic models with a U(3) basis.
In particular, for a binary configuration it is the (modified) vibron model
of U(4) dynamical algebra
\cite{vibron},
which is a group-theoretical model of the dipole motion.
(The modification means a truncation of the basis due to the 
Pauli-principle
\cite{sacm}.)
For a ternary configuration the two independent Jacobi-coordinates
are described by the U(7) model 
\cite{bipr,biap,u7ha,u7di}.
% apart from the Pauli-forbidden states.
For a three-cluster configuration this model has a group-structure of
U$^{ST}_{C_1}$(4)$\otimes$U$_{C_1}$(3) $\otimes$
U$^{ST}_{C_2}$(4)$\otimes$U$_{C_2}$(3) $\otimes$
U$^{ST}_{C_3}$(4)$\otimes$U$_{C_3}$(3) $\otimes$
U$_R$(7).
The exclusion of the Pauli-forbidden states requires the truncation
of the basis of the U(7) model, that determines the lowest allowed major shell,
and in addition 
one needs to distinguish between the Pauli-allowed and forbidden
states within a major shell, too. Different methods can be applied
to this purpose; e.g. one can make an intersection with the U(3) shell
model basis of the nucleus, which is constructed to be free from
the forbidden states.

The SACM is fully algebraic, and semimicroscopic in the sense
discussed above.

When we are interested only in spin-isospin zero states of the nucleus
(a typical problem in cluster studies, and being our case here, too),
then only the space symmetries are relevant (apart from the construction
of the model space).
For a ternary cluster configuration
the corresponding group-chain is
%Eq.2    
\begin{eqnarray}
&&U_{C_1}(3) \otimes U_{C_2}(3) \otimes U_{C_3}(3) \otimes U_R(7) \supset 
\nonumber\\
&&U_C(3) \otimes \{ U_R(6) \supset U_R(3) \} \supset 
\nonumber\\   
&&U(3) \supset SU(3) \supset SO(3) \supset SO(2).
\end{eqnarray}
The basis defined by this chain is especially useful
for treating the exclusion principle, since the U(3) generators commute
with those of the permutation group, therefore, all the basis states of an 
irrep are either Pauli-allowed, or forbidden
\cite{horisup}.           
By applying basis (2) we can  pick up the allowed cluster
states from the U(3) shell model basis (1).         

A Hamiltonian corresponding to the dynamical symmetry of group-chain (2) reads as
%Eq.3     
\begin{eqnarray}
{\hat H} &=&
{\hat H_{C_1}} + {\hat H_{C_2}} + {\hat H_{C_3}} + {\hat H_{U_R(7)}}
\nonumber\\
&+&
{\hat H_{U_C(3)}} + {\hat H_{U_R(6)}}  + {\hat H_{U_R(3)}}
\nonumber\\
&+&
{\hat H_{U(3)}} + {\hat H_{SU(3)}} + {\hat H_{SO(3)}}.
\end{eqnarray}
We note here that the first part 
%Eq.4
\begin{eqnarray}
{\hat H_{CM}} &=&
{\hat H_{C_1}} + {\hat H_{C_2}} + {\hat H_{C_3}} + {\hat H_{U_R(7)}} 
\nonumber\\
&+&
{\hat H_{U_C(3)}} + {\hat H_{U_R(6)}} + {\hat H_{U_R(3)}} 
\end{eqnarray}
is an operator that corresponds to  the pure cluster picture, while the second part
%Eq.5
\begin{equation}
{\hat H_{SM}} =
{\hat H_{U(3)}} + {\hat H_{SU(3)}} + {\hat H_{SO(3)}} 
\end{equation}
is a shell model Hamiltonian (of the united nucleus). 
\\

\noindent
{\it 
The multichannel dynamical symmetry}
(MUSY)
\cite{mus1,mus2} 
connects different cluster configurations (including the shell model limit)
in a nucleus.
Here the word channel refers to the reaction channel
that defines the cluster configuration.

The simplest case is a two-channel symmetry connecting two different
clusterizations. It holds when both cluster configurations can be described
by a U(3) dynamical symmetry, and in addition a further symmetry connects 
them to each other. This latter symmetry acts in the
pseudo space of the particle indices
\cite{hori,mus2}. 
The ${\hat H_{SM}}$ Hamiltonian of Eq. (5)
is symmetric with respect to these transformations, therefore, it is invariant
under the changes from one clusterization to the other.
The cluster part of the Hamiltonian, 
${\hat H_{CM}}$
is affected by the transformation from one configuration to the other,
of course. Nevertheless, it may remain invariant, which is the case for
simple operators, like the harmonic oscillator Hamiltonian, or
the quadrupole operator
\cite{mus2}.
Due to this symmetry of the quadrupole operator, the $E2$ transitions
of different clusterizations also coincide, when the MUSY holds,
just like the energy eigenvalues of the symmetric Hamiltonians
\cite{mus2}.

The MUSY is a composite symmetry of a composite system.
Its logical structure is somewhat similar to that of the dynamical supersymmetry
(SUSY) of nuclear spectroscopy 
\cite{susy}.
In the SUSY case the system has two
components, a bosonic and a fermionic one, each of them showing a
dynamical symmetry, and a further symmetry connects them to each other.
The connecting symmetry is that of the supertransformations 
which change  bosons into  fermions, or vice versa. 
In the MUSY case the system has two (or more) different clusterizations,
each of them having  dynamical symmetries which are connected to each other
by the symmetry of the
pseudo space of the particle indices that change
from one configuration to the other.     
               
When the multichannel dynamical symmetry holds then the spectra
of different clusterizations are related to each other by very strong constraints.
The MUSY provides us with a unified multiplet structure of different cluster configurations,
furthermore the corresponding energies and $E2$
transitions coincide exactly. 
Of course, it can not be decided a priori whether the MUSY holds or not,
rather one can suppose the symmetry and compare its consequences
with the experimental data.

\begin{figure}[placement !]
\includegraphics[height=8.5cm,angle=0.]{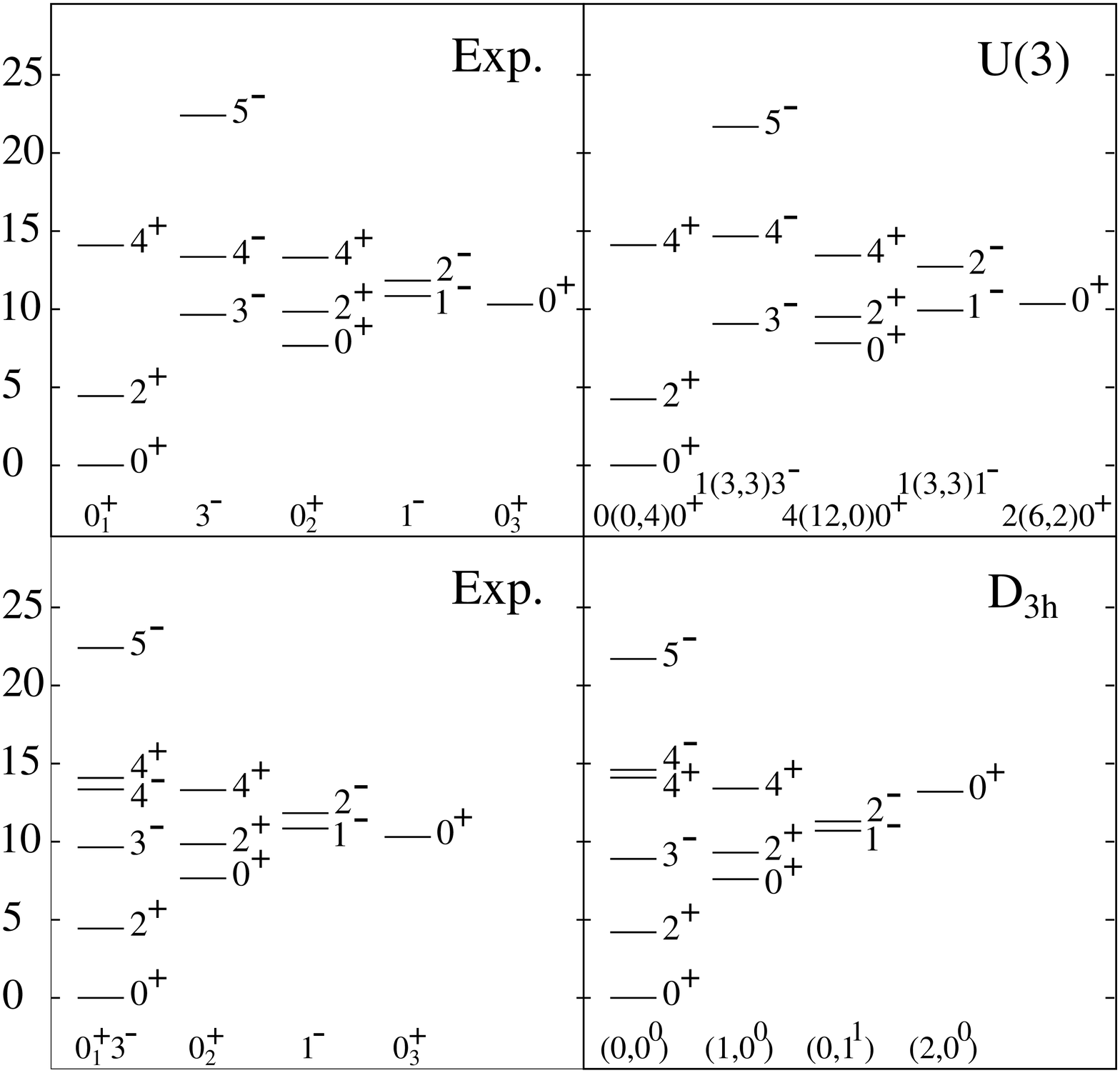}
\caption{Comparison between the experimental and model spectra
of the  $^{12}$C nucleus.
The energy scale is given in MeV, and the bands of the experimental
part are labelled by the $K^{\pi}$ values. The $D_{3h}$ (as well as the experimental 
spectrum) is taken from 
\cite{evid}, and the model bands are labelled by the 
rotational-vibrational quantum numbers of the $D_{3h}$ model.
In the  U(3) MUSY spectrum the bands have the  
 $n (\lambda, \mu)K^{\pi}$ quantum numbers.
\label{fig:spectrum}}
\end{figure}

The energy spectrum of Figure 1 was calculated from the formula
%Eq.6
\begin{eqnarray}
{E} &=&
e\:n  + a\:(\lambda^2 + \mu^2 + \lambda \mu + 3 \lambda + 3 \mu) 
\nonumber
\\
&+& b\:(\lambda - \mu)(\lambda + 2 \mu + 3)(2 \lambda + \mu + 3)
\nonumber
\\
&+& f K^2 + d\:{1 \over {2\theta}}{L(L+1)}.
\end{eqnarray}
In the first term $n$ is the number of  oscillator quanta.
The second term is the expectation value of the second-order 
invariant operator of the SU(3) algebra, which represents
quadrupole-quadrupole interaction.
The third one is the eigenvalue of the third-order invariant
distinguishing between the prolate and oblate shapes.
The $K$-dependent term splits the bands belonging to the same
SU(3) representation. (The corresponding operator is determined
by the operators of the integrity basis of the SU(3) algebra, and is very 
nearly diagonal in the SU(3) basis states
\cite{kdep}.)
In the last part 
$\theta$ is the moment of inertia calculated classically for the rigid shape
determined by the U(3) quantum numbers (for a rotor with axial symmetry)
\cite{saqm}.
The parameters were fitted to the experimental data:
$e = 37.8$ MeV,
$a =-2.4879 $ MeV,
$b =0.045432 $ MeV,
$f =-0.98366 $ MeV,
$d =  1.1661$ MeV.

As it is discussed above, the U(3) MUSY connects the cluster and
quartet (i.e. shell) descriptions. Therefore, in determining its parameters
some shell-model constraints (e.g. systematics) can be, and 
in some cases has been applied
\cite{csri}. In the present study, however, we determined
the parameters from the experimental spectrum, like in
the work
\cite{evid}, in order to treat the two descriptions 
(based on the $D_{3h}$ and U(3) MUSY) on an equal footing.
For comparison we show in the lower part of Figure 1 the
result of the ACM, too from
\cite{evid}.
The number of parameters is comparable in the two
cases: 5 in our Eq. (6), and 6 in Eq. (2) of 
\cite{evid}.

In comparing our semimicroscopic description to that of the
no-core symplectic shell model
\cite{luis}, it is worth noting that the 
lowest-grade SU(3) symmetries we associate with the ground-state band 
and to the Hoyle-band  are the dominating ones in the
fully microscopic description, too. In particular, the (0,4) basis has the 
largest contribution to the ground state, 
and (12,0) symmetry is the head of the symplectic band, dominating the 
Hoyle-states.

\noindent
{\it 
To sum up:}
In Figure 1 we compared the model spectra of two algebraic descriptions to
the experimental spectrum of the  $^{12}$C nucleus.
The result illustrates both the usefulness of the U(7) dynamical algebra in
the treatment of the three-cluster problem,  and the fact that it incorporates
different models. In particular, the algebraic cluster model with $D_{3h}$
interaction and the semimicroscopic algebraic cluster model with the U(3)
multichannel dynamical symmetry give very similar descriptions.
Therefore, further experimental details (on the ``missing states''
and on the transitions) seem to be essential in order to decide which symmetry 
is realized to  a better approximation.  
(At the same time they can  deepen our understanding
in terms of the fully microscopic theories.) From the viewpoint of the 
symmetry studies the combination of the two algebraic methods
can also be very informative:
the operators with the $D_{3h}$ symmetry of the ACM
could be applied on the model  space of the SACM,
which incorporates the Pauli-principle. 
This algebraic treatment would include all the symmetries
applied only partly so far by different models (Table 1).

%\section*{Acknowledgment}
\noindent
{\it This work was supported  by the Hungarian Scientific 
Research Fund -  OTKA (Grant No. K106035).
Helpful discussions on this topic  with 
M. Freer,
F. Iachello,
K. Kat\=o,
G. L\'evai,
Tz. Kokalova, and
A. Merchant,
as well as the technical help by G. Riczu are kindly acknowledged.
The encouragement of the Birmingham group for 
preparing this publication is also highly appreciated.
}

\end{document}